\documentclass{article}
\usepackage{spconf,amsmath,graphicx}

\usepackage{multirow}
\usepackage{lscape}
\usepackage{xcolor}
\usepackage{url,graphicx,amsmath,graphicx,multirow,commath,cite,enumitem,amssymb,hhline}

\title{The Multi-speaker multi-style voice cloning challenge 2021}

\name{\fontsize{9.0pt}{\baselineskip}\selectfont Qicong Xie$^1$, Xiaohai Tian$^2$, Guanghou Liu$^1$, Kun Song$^1$, Lei Xie$^{1*}$, Zhiyong Wu$^{3}$, Hai Li$^4$, Song Shi$^4$, Haizhou Li$^{2,5}$, Fen Hong$^6$, Hui Bu$^7$, Xin Xu$^7$\thanks{
*Corresponding author.
The M2VOC challenge website: http://challenge.ai.iqiyi.com/M2VoC}}

\address{
  $^1$ASLP@NPU, School of Computer Science, Northwestern Polytechnical University, China\\
  $^2$Department of Electrical and Computer Engineering, National University of Singapore\\
  $^3$Tsinghua Shenzhen International Graduate School, Tsinghua University, Shenzhen, China\\
  $^4$iQIYI Inc, China,\\
  $^5$Machine Listening Lab, University of Bremen, Germany\\
  $^6$Originbeat Inc, China\\
  $^7$Beijing Shell Shell Technology Co., Ltd, China\\
\fontsize{9.0pt}{\baselineskip}\selectfont xieqicong@mail.nwpu.edu.cn, lxie@nwpu.edu.cn, marketing@originbeat.com}

\begin{document}
\topmargin=0mm

\ninept

\maketitle

\begin{abstract}
The Multi-speaker Multi-style Voice Cloning Challenge (M2VoC) aims to provide a common sizable dataset as well as a fair testbed for the benchmarking of the popular voice cloning task.
Specifically, we formulate the challenge to adapt an average TTS model to the stylistic target voice  with limited data from target speaker, evaluated by speaker identity and style similarity.
The challenge consists of two tracks, namely few-shot track and one-shot track, where the participants are required to clone multiple target voices with 100 and 5 samples respectively. 
There are also two sub-tracks in each track. 
For sub-track a, to fairly compare different strategies, the participants are allowed to use only the training data provided by the organizer strictly.
For sub-track b, the participants are allowed to use any data publicly available.
In this paper, we present a detailed explanation on the tasks and data used in the challenge, followed by a summary of submitted systems and evaluation results.

\end{abstract}
%

\vspace{-0.3cm}
\section{Introduction}
\label{sec:intro}
\vspace{-0.1cm}
With the development of deep learning, speech synthesis has made significant progress in recent years. While recently proposed end-to-end speech synthesis systems, e.g., Tacotron~\cite{wang2017tacotron}, DurIAN~\cite{Yu2020durian} and FastSpeech~\cite{Ren2019fastspeech}, are able to generate high-quality and natural sounding speech, these models usually rely on a large amount of training data from a single speaker. The speech quality, speaker similarity, expressiveness and robustness of synthetic speech are still not systematically examined for different speakers and various speaking styles, especially in real-world low-resourced conditions, e.g., each speaker only has a few samples at hand for cloning. However, this so-called multi-speaker multi-style voice cloning task has found significant applications on customized TTS. 

Imitating speaking style is one of the desired abilities of a TTS system. Several strategies have been recently proposed to model stylistic or expressive speech for end-to-end TTS. Speaking style comes with different patterns in prosody, such as rhythm, pause, intonation, and stress, etc. Hence direct modeling prosodic aspects of speech is beneficial for stylization~\cite{lancucki2020fastpitch, raitio2020controllable}. Variational Autoencoder (VAE)~\cite{kingma2013auto,akuzawa2018expressive} and GST~\cite{Wang2018GST} are two typical models built upon sequence-to-sequence (seq2seq) models for style modeling. Global Style Tokens (GSTs)~\cite{Wang2018GST} is introduced for modeling style in an unsupervised way, using multi-head attention to learn a similarity measure between the reference embedding and each token in a bank of randomly initialized embeddings, and the Text-Predicted Global Style Token (TP-GST)~\cite{stanton2018predicting} learns to predict stylistic renderings from text alone, requiring neither explicit labels during training nor auxiliary inputs for inference. Note that these studies modeling speaker styles are mostly based on a large amount of data.

Besides, building a target voice with limited data, i.e., a few minutes or even only several recorded samples, or \textit{voice cloning}, has been a popular research topic recently. Speaker adaptation~\cite{taigman2017voiceloop, chen2018sample} is a straightforward solution which transfers an average voice model to the target voice using the samples of the target speaker while the average voice model is usually built on speech from multiple speakers.
To obtain reasonable speaker voice information, it is very important to accurately obtain speaker identity information.
Disentanglement is a popular solution, and its main idea is to separate the speaker's voice and content information. In\cite{wang2020spoken}, phone posteriorgram (PPG) is used as an intermediate variable to factorize spoken content and voice~\cite{wang2020spoken}. VQ technology~\cite{oord2017neural, razavi2019generating} and U-net structure~\cite{ronneberger2015u} are used to disentangle the speaker's content and voice information for oneshot-VC~\cite{wu2020vqvc+}. 
Speaker embedding is also a reasonable way to model the speaker identity. It can be obtained through a well-designed encoder. In~\cite{choi2020attentron}, a fine-grained encoder extracts variable-length style information via attention mechanism and a coarse-grained encoder greatly stabilizes the speech synthesis to improve robustness. Also, speaker embedding can be obtained from the joint training of a speaker verification model and an acoustic model~\cite{wang2020bi, shi2020aishell, cai2020speaker}. Besides, in real-world voice cloning, low quality recording or noisy speech is another issue to address~\cite{hsu2019disentangling, cong2020data}.

To provide a fair platform for benchmarking the voice cloning task, we launched the Multi-Speaker Multi-Style Voice Cloning Challenge (M2VoC) at ICASSP 2021.
The focus of the challenge is on voice cloning in limited data condition, e.g. 100 and 5 samples from target speaker, which are referred to as few-shot and one-shot tracks, respectively. 
In this challenge, a multi-speaker training data set is provided as a fair test platform comparing different methods. In this paper, we present a summary of the challenge, including various methods provided by participants and the corresponding results.  Section~\ref{sec:pagestyle} describes the details of the released data and the participation guide. Section~\ref{sec:typestyle} summarizes the systems submitted by the participants. Section~\ref{sec:majhead} introduces the evaluation method and results. Finally, conclusions are drawn in Section~\ref{sec:print}.
\vspace{-0.3cm}
\section{Data and tracks}
\label{sec:pagestyle}
\vspace{-0.2cm}

\subsection{Data}
\vspace{-0.2cm}

We provide four audio/text data sets to the participants at different stages of the challenge. All audio data is mono, 44.1KHz sampling rate, 16 bits, equipped with transcripts. The language is Mandarin.

\textbf{Multi-speaker training set (MST):} This part of data consists of two subsets, including the AIShell-3~\cite{Shi2020} data set, called MST-AIShell in the challenge. The data set contains about 85 hours of Mandarin speech data from 218 common speakers, which is recorded through a high-fidelity microphone in an ordinary room with some inevitable reverberation and background noise. Another data set, called MST-Originbeat, consists of two professional Mandarin speakers (one male and one female), and the high-quality voice data (5,000 utterances per speaker) is recorded with a high-fidelity microphone in a recording studio without obvious reverberation and background noise. Text annotations are provided with Pinyin and prosodic boundary labels.

\textbf{Target speaker validation set (TSV):} For each track, there are two validation target speakers with different speaking styles.
For track 1, each speaker has 100 speech samples, and for track 2, each speaker has 5 speech samples.
The voice data is recorded using a high-fidelity microphone in a recording studio. Text annotations are provided with Pinyin and prosodic boundary labels. These two validation target speakers are provided to the participants for voice cloning experiments during the system building stage.

\textbf{Target speaker test set (TST):} For each track, three target speakers with different speaking styles (different from those in TSV) are released for testing and ranking.
Again, for track 1, each speaker has 100 speech samples, and for track 2, each speaker has 5 speech samples.
Voice data is recorded using a high-fidelity microphone in a recording studio. Text annotations are provided with Pinyin and prosodic boundary labels.

\textbf{Test text set (TT):} This text set includes the sentences (with Pinyin annotations) provided to the participants for speech synthesis for the test speakers in TST.
The sentences can be divided into three categories, namely, style, common, and intelligibility. The sentences in the style set are in-domain sentences for style imitation test. That is to say, the sentences match the same text domain in style of the speaker in TST. The sentences in the common set are out-of-domain sentences (without clear style) and are used for quality and similarity tests. The intelligibility test set includes non-sense sentences. For track 1, participants are asked to synthesize speech for the three TST speakers on the style, common and intelligibility sets; for track 2, participants are asked to synthesize speech for the three TST speakers on the common and intelligibility sets.

\vspace{-0.3cm}

\subsection{Tracks}
\vspace{-0.2cm}

\qquad
\vspace{-0.3cm}

\textbf{Track 1 (Few-shot track):} The organizers provide two and three target speakers for voice cloning validation (TSV) and evaluation (TST) respectively. Each speaker has different speaking style with 100 available samples. The organizers also provide a large multi-speaker dataset (MST) for participants, which can be used as training data for a base model. The test text set (TT) is a list of sentences for TTS generation for the three target speakers in TST.

\textbf{Track 2 (One-shot track):} For track 2, requirements are the same as track 1, except that only 5 recordings are provided for each target speaker. In this track, participants only need to synthesize speech on the common and intelligibility sets. 

For both tracks, we also set up two subtracks:
\vspace{-0.2cm}

\begin{itemize}
    \item Sub-track a: The building of the TTS system is strictly limited to the released data by the organizer. Using extra data is prohibited.
    \vspace{-0.1cm}
    \item Sub-track b: Besides the released data, the participants can use any data set publicly available for system building. However, the participants have to clearly list the data used in the final system description paper.
\end{itemize}
\vspace{-0.25cm}

\vspace{-0.2cm}

\section{Participants and submitted systems}
\label{sec:typestyle}
\vspace{-0.1cm}

In total, we collected submissions from 26 teams, including 2 baselines (B01 and B02) and 24 participants (T01 to T24) from both academia and industry. 
Specifically, 18 and 22 teams submitted their results to Track 1a and Track 1b respectively, while 17 and 19 teams submitted their results to Track 2a and Track 2b respectively.  
It should be noted that 9 teams did not submit an appropriate system description. So we exclude them in the following discussion.
Here we summarize the approaches used by the top-ranking submissions, according to their system descriptions.

\vspace{-0.3cm}

\subsection{Acoustic model}
\label{subs:am}
\vspace{-0.2cm}

According to the submitted system descriptions, the acoustic models can be grouped into two categories: 1) autoregressive (AR) model and 2) non-autoregressive (Non-AR) model. Note that most teams use the same type of acoustic model for both Track 1 and Track 2, while T20 chose different solutions. 

In the AR acoustic model category, the input phoneme sequence is first encoded by the encoder. Then the decoder generates the target spectral features in an autoregressive manner.
In the submissions, Tacotron~\cite{wang2017tacotron, Shen2018taco2} is the most popular one, where an encoder-attention-decoder based architecture is adopted for autoregressive generation. Duration Informed Attention Network (DurIAN)~\cite{Yu2020durian} is another popular AR model in which the alignments between the input text and the output acoustic features are inferred from a duration model with the added benefit that the duration predictor is significantly resilient to failures that disturb the attention mechanism. Non-Attentive Tacotron~\cite{Shen2020nattaco} is similar to DurIAN, which introduces Gaussian upsampling method significantly improving the naturalness compared to the vanilla upsampling through repetition.


A typical representative of non-AR model used by the participants is FastSpeech~\cite{Ren2019fastspeech}, where the duration information is extracted from a pre-trained teacher forcing AR model. Then the text content and the duration are used as the input to predict the target spectral features in a non-autoregressive mode. 
Extended from FastSpeech, FastSpeech2~\cite{Ren2020fs2} replaces the teacher-student distillation with an external aligner. Moreover, FastSpeech2 introduces more variance information (e.g., duration, pitch, energy, etc.) to ease the one-to-many mapping problem in speech prediction.

\vspace{-0.3cm}

\subsection{Vocoder}
\label{subs:vo}
\vspace{-0.2cm}


Similarly, the vocoders used in the submitted systems can be divided into autoregressive and non-autoregressive as well. Specifically, 5 and 10 teams chose the AR and non-AR neural vocoders respectively. 
Among the AR vocoders, 4 teams used LPCNet~\cite{Valin2019lpcnet} and one team used SC-WaveRNN~\cite{paul2020speaker} in their implementations. For the submissions using non-AR vocoders,  HiFi-GAN~\cite{su2020hifi} was adopted in 4 submissions, while MelGAN~\cite{kumar2019melgan, yang2020multi}, Parallel WaveGAN~\cite{yamamoto2020parallel} and WaveGrad~\cite{chen2020wavegrad} were also used in some other implementations.

\vspace{-0.3cm}

\subsection{Speaker and style modeling}
\label{subs:sv}
\vspace{-0.2cm}

The speaker and style imitation is the key of multi-speaker multi-style voice cloning. Robust speaker and style representations are crucial to model and generate the target voice with desired speaker identity and style. According to the submitted system descriptions, the speaker and style control method can be grouped into two categories: 1) Using a single speaker representation for both speaker and style modeling and 2) controlling the speaker and style with different methods.



\textbf{Speaker representation:} In the submitted systems, using speaker embedding for both speaker identity and style modeling is one of the main streams. In general, the speaker embedding can be obtained by either pretrained with a separate speaker model or jointly-optimized with the acoustic model. 
In some works, a pretrained network was utilized to extract the speaker embedding. In details, a separate neural network model is first trained to perform the speaker verification (SV) task. Then the speaker embedding is extracted from the middle layer of the model as a conditional input representing the speaker identity for the multi-speaker TTS training. Various SV models were used in this challenge. D-vector~\cite{Wan2018dvector}, ResNet34~\cite{2020Onresnet34} and x-vector~\cite{2018Xxvector} were used in T07, T16 and T23 respectively.
Speaker embedding can also be obtained from a voice conversion (VC) system.
In T14~\cite{T142021M2VoC}, the target speaker encoder in AdaIN-VC~\cite{Chou2019VC} was adopted to extract speaker information from the target utterance. AdaIN-VC~\cite{Chou2019VC} is a one-shot VC system that shows good generalization ability for unseen speakers.
For the joint optimization solution, the straightforward method is speaker look-up table, in which each speaker code is encoded in a trainable embedding table, jointly trained with acoustic model via backpropagation. T06 and T24~\cite{T242021M2VoC} have used this solution in their systems. 

\textbf{Style modeling:} Alternative to the solution of using a single representation for both speaker and style modeling, some explicit style modeling approaches are proposed to improve the speaking style imitation of the generated voice.
For example, T03~\cite{T032021M2VoC} has proposed a prosody and voice factorization strategy for few-shot speaker adaption. Specifically, a set of auto-learned phoneme-level transition tokens were used to help the attention control the duration of each phoneme. T23 proposed an adversarially-trained speaker classifier to prevent the text encoding from capturing speaker information. T18 proposed to use a BERT~\cite{devlin2018bert} module to predict the break of each Chinese character in an input sentence. T15~\cite{T152021M2VoC} used a fine-grained encoder added at the decoder’s tailor, which extracts variable-length detailed style information from multiple reference samples via an attention mechanism. T03 and T15 also used global style tokens (GST) for both speaker and style control, which consists of a reference encoder, style attention, and style embedding. This module can compress a variable-length audio signal into a fixed-length vector, which is considered to be related to the speaker's identity and style. The GST module is jointly trained with the rest of the model, driven only by the reconstruction loss.

\vspace{-0.3cm}

\subsection{Speaker and style adaptation}
\label{adapt}
\vspace{-0.2cm}

Adapting a multi-speaker average voice model to the target speaker is commonly used in the submissions. The base voice model is built with the released multi-speaker data (MST) for sub-track a and other public data for sub-track b. In general, there are mainly two methods utilized in the submitted systems to adapt the average model towards the target speaker in terms of the speaker identity and style.

\textbf{Embedding based model adaptation:} In this method, a embedding vector or embedding network is used to condition the speaker identity and style of the training voice. During training the average model, the embedding vector or network is used to distinguish the acoustic features from different speaker identities and speaking styles. At run-time, the target speaker representations are utilized as a part of input feature to generate the voice of target speaker.

\textbf{Fine-tuning based model adaptation:} This method first uses multi-speaker data to train an average model. For a given target speaker, the average model is then adjusted with the small amount of target data.  The adaptation can be achieved by fine-tuning either all the parameters of the model or a part of them. According to the submitted systems, T11 selects to freeze some model parameters and fine-tunes other parameters of the model, while T13~\cite{T132021M2VoC} selects to fine-tune all parameters of the model.

\vspace{-0.2cm}

\section{Subjective evaluation}
\label{sec:majhead}
\vspace{-0.1cm}

The evaluation and ranking adopted for M2VoC are based on subjective listening test. The experimental design and results are presented in this section.
\vspace{-0.3cm}

\subsection{Evaluation methodology}
\vspace{-0.2cm}

We conducted mean opinion score (MOS) tests to assess speech quality, speaker similarity and style similarity of the generated voice from different submissions.

\begin{itemize}[leftmargin=*]
\itemsep0em

\item {\textbf{Speech quality}:} In each session, listeners listened to one sample and chose a score which represented how natural or unnatural the sentence sounded on a scale of 1 [Completely unnatural] to 5 [Completely natural].

\item {\textbf{Speaker similarity}:}
In each session, listeners could play 2 reference samples of the original speaker and one synthetic sample. They chose a response that represented how similar the speaker identity of the synthetic speech sounded to that of the voice in the reference samples on a scale from 1 [Sounds like a totally different person] to 5 [Sounds like exactly the same person].

\item {\textbf{Style similarity}:}
In each session, listeners could play 2 reference samples of the original speaker and one synthetic sample. They chose a response that represented how similar the speaking style of the synthetic speech sounded to that of the voice in the reference samples on a scale from 1 [Sounds like a totally different style] to 5 [Sounds like exactly the same style].
\end{itemize}

Note that speech quality and speaker similarity were tested on both the common and style sets. While the style similarity was only tested on the style set. We took two rounds of subjective evaluation: the first round included submissions from all teams and the second round included several top-scoring teams (with no significant score differences from the 2nd place team of the first round) for further evaluation. The final winner for each track was selected based on the combined results from both rounds. There were 66 and 30 listeners participated the first and second rounds of the subjective listening tests, respectively. 
All the listeners are native Chinese -- a combination of college students in linguistics major and professional speech annotators. 

Considering the huge cost of subjective evaluation on the quality, style and similarity in a short period, we adopted a sampling evaluation method. 
In the first round of subjective evaluation, for each test, every listener received 10 sentences randomly selected from 100 generated samples for each test set of each speaker. 
In the second round, the number of evaluation sentences became 20, while the other settings were the same as the first round.
The three sets of subjective evaluations were done separately, that is to say, each sentence was evaluated three times.
A total of 96 listeners (9 male and 87 female) completed the evaluation, for a total of 181,000 and 66,600 evaluated utterances in first and second round, respectively.

The final result is the average MOS score of quality, style, and similarity for track 1 and quality and similarity for track 2. Note that we did not consider the intelligibility test for ranking this time due to the tight evaluation period. However, we found that most submitted systems can generate reasonably intelligible speech.


\begin{table}[t]
\centering
\caption{MOS of submitted systems on Track 1.}
\setlength{\tabcolsep}{1.0mm}
\small
\label{tb:ct1a}
\begin{tabular}{llll}
\hline
\multicolumn{1}{c}{Team ID} & \multicolumn{1}{c}{Quality} & \multicolumn{1}{c}{Similarity} & \multicolumn{1}{c}{Style} \\ \hline
\multicolumn{4}{c}{Track 1a}                               \\ \hline
T22     & 4.2373±(0.0295) & 4.2484±(0.0247) & 4.1488±(0.0227) \\
T15     & 4.0214±(0.0339) & 4.1370±(0.0270) & 4.1212±(0.0227) \\
T03     & 4.1741±(0.0297) & 4.1455±(0.0269) & 3.9348±(0.0262) \\
T13     & 4.0623±(0.0307) & 3.8832±(0.0314) & 3.8027±(0.0261) \\ \hline
\multicolumn{4}{c}{Track 1b}                               \\ \hline
T22 & 4.3132±(0.0275) & 4.2466±(0.0249) & 4.1056±(0.0230) \\
T03 & 4.2636±(0.0283) & 4.1027±(0.0285) & 3.8815±(0.0272) \\
T19 & 4.0502±(0.0339) & 4.0305±(0.0290) & 3.9574±(0.0250) \\
T18 & 4.1877±(0.0282) & 3.8673±(0.0341) & 3.8926±(0.0256) \\
T24 & 3.9502±(0.0358) & 3.9861±(0.0304) & 3.8997±(0.0248) \\
T06 & 3.8477±(0.0384) & 3.9564±(0.0327) & 3.9438±(0.0255) \\
T13 & 4.1941±(0.0272) & 3.8355±(0.0331) & 3.7517±(0.0272) \\  \hline
\end{tabular}
\vspace{-0.5cm}
\end{table}


\begin{table}[t]
\centering
\caption{MOS of submitted systems on Track 2.}
\setlength{\tabcolsep}{5.1mm}
\small
\label{tb:ct2a}
\begin{tabular}{lll}
\hline
\multicolumn{1}{c}{Team ID} & \multicolumn{1}{c}{Quality} & \multicolumn{1}{c}{Similarity} \\ \hline
\multicolumn{3}{c}{Track 2a}                               \\ \hline
T03 & 4.0905±(0.0451) & 3.2168±(0.0605) \\
T14 & 3.8941±(0.0476) & 3.2205±(0.0595) \\
T10 & 3.8768±(0.0499) & 3.2250±(0.0612) \\
T15 & 3.9568±(0.0493) & 3.1300±(0.0604) \\
T18 & 3.8700±(0.0479) & 3.1368±(0.0598) \\
T24 & 3.4900±(0.0565) & 3.0895±(0.0620) \\ \hline
\multicolumn{3}{c}{Track 2b}                               \\ \hline
T18 & 4.1650±(0.0399) & 3.2409±(0.0635) \\
T03 & 4.1086±(0.0450) & 3.1964±(0.0645) \\
T14 & 3.8718±(0.0519) & 3.1445±(0.0634) \\
T24 & 3.4818±(0.0573) & 3.1445±(0.0634) \\ \hline
\end{tabular}
\vspace{-0.5cm}
\end{table}
\vspace{-0.3cm}

\subsection{Evaluation results}
\vspace{-0.2cm}

\begin{figure}[t]
    \centering
    \includegraphics[scale=0.63]{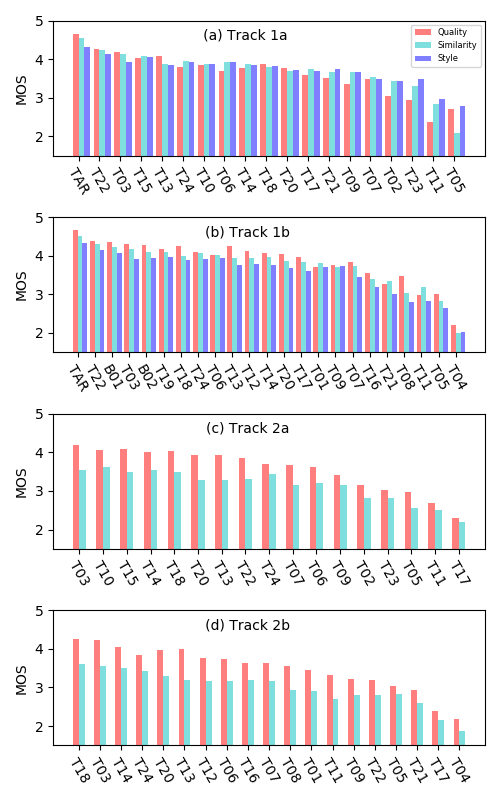}
    \caption{MOS of all submissions in the first round evaluation on each track.}
    \label{fig:my_label}\vspace{-0.6cm}
\end{figure}

Fig.~\ref{fig:my_label} shows the results of all participating teams in the first round evaluation. TAR refers to the natural speech of the target speaker.
Table \ref{tb:ct1a} and \ref{tb:ct2a} are the combined results of the two rounds for the top-performing teams in each track.

\textbf{Track 1:} For Track 1a, the top 4 teams have done a good job in all aspects, with MOS scores around 4, especially T22 which has achieved the highest scores in the three testing metrics. It can be seen that the current few-shot task (with around 100 training samples) can obtain a reasonable performance through, for example, FastSpeech+HiFiGAN or Tacotron+LPCNet.
Compared with Track 1a, the participants can use additional training data on Track 1b. Comparing the two sub-tracks, the performance on sound quality of most top teams has improved on Track 1b. This shows that additional data is beneficial to sound quality.
However, in terms of style and speaker similarity, the performance of most top teams on Track 1b is not significantly better than that on Track 1a.
This may suggest that the additional data may not be of much use for modeling speaker and style under current challenge data setup.

\textbf{Track 2:} For Track 2a, the quality of synthetic speech is acceptable in general for the top 7 teams, and the result of sub-track 2b is better than that of sub-track 2a, indicating that increasing the amount of training data is helpful to the sound quality.
Among the participating teams, T03 and T18 have achieved first place in the two sub-tracks respectively. The two teams both used LPCNet as the neural vocoder, which indicates that LPCNet is robust in waveform generation and may have great potential to synthesize high-quality audio in low-resource application scenarios.
For similarity, the result of Track 2 is very different from that of Track 1, and the MOS scores for the top teams are only over 3. Problems of unstable pronunciation and poor speaker similarity can be easily found from the synthetic speech.
Compared with sub-track 2a, the performance improvement on sub-track 2b is almost negligible, which shows that on one-shot task, the additional training data may not be useful for speaker voice cloning.
The task of voice cloning with only several recorded samples is still very challenging.

\vspace{-0.3cm}

\section{Conclusion}
\label{sec:print}
\vspace{-0.2cm}

This paper summarizes the multi-speaker and multi-style voice cloning challenge held in ICASSP 2021. The M2VoC challenge aims to provide a common data set and a fair test platform for research on voice cloning tasks. The challenge demonstrated the performance of current voice cloning technologies -- with the advances of deep learning, few-shot voice cloning has achieved reasonably good performance but one-shot voice cloning is still an unsolved problem. Note that in real-world voice cloning applications, low-quality (noisy) audio and time/cost constrains of training/adaptation/inference are also important factors that can not be ignored. We may consider these factors in the next challenge.

\vfill
\pagebreak


\scriptsize
\bibliographystyle{IEEEbib}
\bibliography{strings,refs}

\begin{thebibliography}{10}

\bibitem{wang2017tacotron}
Yuxuan Wang, RJ~Skerry-Ryan, Daisy Stanton, Yonghui Wu, Ron~J Weiss, Navdeep
  Jaitly, Zongheng Yang, Ying Xiao, Zhifeng Chen, Samy Bengio, et~al.,
\newblock ``Tacotron: Towards end-to-end speech synthesis,''
\newblock {\em arXiv preprint arXiv:1703.10135}, 2017.

\bibitem{Yu2020durian}
Chengzhu Yu, Heng Lu, Na~Hu, Meng Yu, Chao Weng, Kun Xu, Peng Liu, Deyi Tuo,
  Shiyin Kang, Guangzhi Lei, Dan Su, and Dong Yu,
\newblock ``{DurIAN: Duration informed attention network for speech
  synthesis},''
\newblock {\em Proceedings of the Annual Conference of the International Speech
  Communication Association, INTERSPEECH}, vol. 2020-October, pp. 2027--2031,
  2020.

\bibitem{Ren2019fastspeech}
Yi~Ren, Tao Qin, Yangjun Ruan, Sheng Zhao, Tie~Yan Liu, Xu~Tan, and Zhou Zhao,
\newblock ``{FastSpeech: Fast, robust and controllable text to speech},''
\newblock {\em arXiv}, , no. NeurIPS, 2019.

\bibitem{lancucki2020fastpitch}
Adrian {\L}a{\'n}cucki,
\newblock ``Fastpitch: Parallel text-to-speech with pitch prediction,''
\newblock {\em arXiv preprint arXiv:2006.06873}, 2020.

\bibitem{raitio2020controllable}
Tuomo Raitio, Ramya Rasipuram, and Dan Castellani,
\newblock ``Controllable neural text-to-speech synthesis using intuitive
  prosodic features,''
\newblock {\em arXiv preprint arXiv:2009.06775}, 2020.

\bibitem{kingma2013auto}
Diederik~P Kingma and Max Welling,
\newblock ``Auto-encoding variational bayes,''
\newblock {\em arXiv preprint arXiv:1312.6114}, 2013.

\bibitem{akuzawa2018expressive}
Kei Akuzawa, Yusuke Iwasawa, and Yutaka Matsuo,
\newblock ``Expressive speech synthesis via modeling expressions with
  variational autoencoder,''
\newblock {\em arXiv preprint arXiv:1804.02135}, 2018.

\bibitem{Wang2018GST}
Yuxuan Wang, Daisy Stanton, Yu~Zhang, R.~J. Skerry{-}Ryan, Eric Battenberg,
  Joel Shor, Ying Xiao, Ye~Jia, Fei Ren, and Rif~A. Saurous,
\newblock ``Style tokens: Unsupervised style modeling, control and transfer in
  end-to-end speech synthesis,''
\newblock in {\em Proceedings of the 35th International Conference on Machine
  Learning, {ICML} 2018, Stockholmsm{\"{a}}ssan, Stockholm, Sweden, July 10-15,
  2018}, 2018, pp. 5167--5176.

\bibitem{stanton2018predicting}
Daisy Stanton, Yuxuan Wang, and RJ~Skerry-Ryan,
\newblock ``Predicting expressive speaking style from text in end-to-end speech
  synthesis,''
\newblock in {\em 2018 IEEE Spoken Language Technology Workshop (SLT)}. IEEE,
  2018, pp. 595--602.

\bibitem{taigman2017voiceloop}
Yaniv Taigman, Lior Wolf, Adam Polyak, and Eliya Nachmani,
\newblock ``Voiceloop: Voice fitting and synthesis via a phonological loop,''
\newblock {\em arXiv preprint arXiv:1707.06588}, 2017.

\bibitem{chen2018sample}
Yutian Chen, Yannis Assael, Brendan Shillingford, David Budden, Scott Reed,
  Heiga Zen, Quan Wang, Luis~C Cobo, Andrew Trask, Ben Laurie, et~al.,
\newblock ``Sample efficient adaptive text-to-speech,''
\newblock {\em arXiv preprint arXiv:1809.10460}, 2018.

\bibitem{wang2020spoken}
Tao Wang, Jianhua Tao, Ruibo Fu, Jiangyan Yi, Zhengqi Wen, and Rongxiu Zhong,
\newblock ``Spoken content and voice factorization for few-shot speaker
  adaptation,''
\newblock {\em Proc. Interspeech 2020}, pp. 796--800, 2020.

\bibitem{oord2017neural}
Aaron van~den Oord, Oriol Vinyals, and Koray Kavukcuoglu,
\newblock ``Neural discrete representation learning,''
\newblock {\em arXiv preprint arXiv:1711.00937}, 2017.

\bibitem{razavi2019generating}
Ali Razavi, Aaron van~den Oord, and Oriol Vinyals,
\newblock ``Generating diverse high-fidelity images with vq-vae-2,''
\newblock {\em arXiv preprint arXiv:1906.00446}, 2019.

\bibitem{ronneberger2015u}
Olaf Ronneberger, Philipp Fischer, and Thomas Brox,
\newblock ``U-net: Convolutional networks for biomedical image segmentation,''
\newblock in {\em International Conference on Medical image computing and
  computer-assisted intervention}. Springer, 2015, pp. 234--241.

\bibitem{wu2020vqvc+}
Da-Yi Wu, Yen-Hao Chen, and Hung-Yi Lee,
\newblock ``Vqvc+: One-shot voice conversion by vector quantization and u-net
  architecture,''
\newblock {\em arXiv preprint arXiv:2006.04154}, 2020.

\bibitem{choi2020attentron}
Seungwoo Choi, Seungju Han, Dongyoung Kim, and Sungjoo Ha,
\newblock ``Attentron: Few-shot text-to-speech utilizing attention-based
  variable-length embedding,''
\newblock {\em arXiv preprint arXiv:2005.08484}, 2020.

\bibitem{wang2020bi}
Tao Wang, Jianhua Tao, Ruibo Fu, Jiangyan Yi, Zhengqi Wen, and Chunyu Qiang,
\newblock ``Bi-level speaker supervision for one-shot speech synthesis,''
\newblock {\em Proc. Interspeech 2020}, pp. 3989--3993, 2020.

\bibitem{shi2020aishell}
Yao Shi, Hui Bu, Xin Xu, Shaoji Zhang, and Ming Li,
\newblock ``Aishell-3: A multi-speaker mandarin tts corpus and the baselines,''
\newblock {\em arXiv preprint arXiv:2010.11567}, 2020.

\bibitem{cai2020speaker}
Zexin Cai, Chuxiong Zhang, and Ming Li,
\newblock ``From speaker verification to multispeaker speech synthesis, deep
  transfer with feedback constraint,''
\newblock {\em arXiv preprint arXiv:2005.04587}, 2020.

\bibitem{hsu2019disentangling}
Wei-Ning Hsu, Yu~Zhang, Ron~J Weiss, Yu-An Chung, Yuxuan Wang, Yonghui Wu, and
  James Glass,
\newblock ``Disentangling correlated speaker and noise for speech synthesis via
  data augmentation and adversarial factorization,''
\newblock in {\em ICASSP 2019-2019 IEEE International Conference on Acoustics,
  Speech and Signal Processing (ICASSP)}. IEEE, 2019, pp. 5901--5905.

\bibitem{cong2020data}
Jian Cong, Shan Yang, Lei Xie, Guoqiao Yu, and Guanglu Wan,
\newblock ``Data efficient voice cloning from noisy samples with domain
  adversarial training,''
\newblock {\em arXiv preprint arXiv:2008.04265}, 2020.

\bibitem{Shi2020}
Yao Shi, Hui Bu, Xin Xu, Shaoji Zhang, and Ming Li,
\newblock ``{AISHELL-3: A Multi-speaker Mandarin TTS Corpus and the
  Baselines},''
\newblock {\em arXiv preprint arXiv:2010.11567}, 2020.

\bibitem{Shen2018taco2}
Jonathan Shen, Ruoming Pang, Ron~J. Weiss, Mike Schuster, Navdeep Jaitly,
  Zongheng Yang, Zhifeng Chen, Yu~Zhang, Yuxuan Wang, Rj~Skerrv-Ryan, Rif~A.
  Saurous, Yannis Agiomvrgiannakis, and Yonghui Wu,
\newblock ``{Natural TTS Synthesis by Conditioning Wavenet on MEL Spectrogram
  Predictions},''
\newblock in {\em ICASSP, IEEE International Conference on Acoustics, Speech
  and Signal Processing - Proceedings}, 2018, vol. 2018-April, pp. 4779--4783.

\bibitem{Shen2020nattaco}
Jonathan Shen, Ye~Jia, Yu~Zhang, Isaac Elias, Heiga Zen, Yonghui Wu, and Mike
  Chrzanowski,
\newblock ``{Non-Attentive Tacotron: Robust and controllable neural TTS
  synthesis including unsupervised duration modeling},'' 2020.

\bibitem{Ren2020fs2}
Yi~Ren, Chenxu Hu, Xu~Tan, Tao Qin, Sheng Zhao, Zhou Zhao, and Tie~Yan Liu,
\newblock ``{FastSpeech 2: Fast and High-Quality End-to-End Text to Speech},''
\newblock {\em arXiv}, pp. 1--11, 2020.

\bibitem{Valin2019lpcnet}
Jean~Marc Valin and Jan Skoglund,
\newblock ``{LPCNET: Improving Neural Speech Synthesis through Linear
  Prediction},''
\newblock {\em ICASSP, IEEE International Conference on Acoustics, Speech and
  Signal Processing - Proceedings}, vol. 2019-May, pp. 5891--5895, 2019.

\bibitem{paul2020speaker}
Dipjyoti Paul, Yannis Pantazis, and Yannis Stylianou,
\newblock ``Speaker conditional wavernn: Towards universal neural vocoder for
  unseen speaker and recording conditions,''
\newblock {\em arXiv preprint arXiv:2008.05289}, 2020.

\bibitem{su2020hifi}
Jiaqi Su, Zeyu Jin, and Adam Finkelstein,
\newblock ``Hifi-gan: High-fidelity denoising and dereverberation based on
  speech deep features in adversarial networks,''
\newblock {\em arXiv preprint arXiv:2006.05694}, 2020.

\bibitem{kumar2019melgan}
Kundan Kumar, Rithesh Kumar, Thibault de~Boissiere, Lucas Gestin, Wei~Zhen
  Teoh, Jose Sotelo, Alexandre de~Brebisson, Yoshua Bengio, and Aaron
  Courville,
\newblock ``{MelGAN: Generative adversarial networks for conditional waveform
  synthesis},''
\newblock {\em Advances in Neural Information Processing Systems}, vol. 32, no.
  NeurIPS 2019, 2019.

\bibitem{yang2020multi}
Geng Yang, Shan Yang, Kai Liu, Peng Fang, Wei Chen, and Lei Xie,
\newblock ``Multi-band melgan: Faster waveform generation for high-quality
  text-to-speech,''
\newblock {\em arXiv preprint arXiv:2005.05106}, 2020.

\bibitem{yamamoto2020parallel}
Ryuichi Yamamoto, Eunwoo Song, and Jae-Min Kim,
\newblock ``Parallel wavegan: A fast waveform generation model based on
  generative adversarial networks with multi-resolution spectrogram,''
\newblock in {\em ICASSP 2020-2020 IEEE International Conference on Acoustics,
  Speech and Signal Processing (ICASSP)}. IEEE, 2020, pp. 6199--6203.

\bibitem{chen2020wavegrad}
Nanxin Chen, Yu~Zhang, Heiga Zen, Ron~J Weiss, Mohammad Norouzi, and William
  Chan,
\newblock ``Wavegrad: Estimating gradients for waveform generation,''
\newblock {\em arXiv preprint arXiv:2009.00713}, 2020.

\bibitem{Wan2018dvector}
Li~Wan, Quan Wang, Alan Papir, and Ignacio~Lopez Moreno,
\newblock ``{Generalized end-to-end loss for speaker verification},''
\newblock {\em ICASSP, IEEE International Conference on Acoustics, Speech and
  Signal Processing - Proceedings}, vol. 2018-April, pp. 4879--4883, 2018.

\bibitem{2020Onresnet34}
Weicheng Cai, Jinkun Chen, Jun Zhang, and Ming Li,
\newblock ``On-the-fly data loader and utterance-level aggregation for speaker
  and language recognition,''
\newblock {\em IEEE/ACM Transactions on Audio, Speech, and Language
  Processing}, vol. PP, no. 99, pp. 1--1, 2020.

\bibitem{2018Xxvector}
David Snyder, Daniel Garcia-Romero, Gregory Sell, Daniel Povey, and Sanjeev
  Khudanpur,
\newblock ``X-vectors: Robust dnn embeddings for speaker recognition,''
\newblock in {\em ICASSP 2018 - 2018 IEEE International Conference on
  Acoustics, Speech and Signal Processing (ICASSP)}, 2018.

\bibitem{T142021M2VoC}
Chungming Chien, Jhenghao Lin, Chienyu Huang, Pochun Hsu, and Hungyi Lee,
\newblock ``Investigating on incorporating pretrained and learnable speaker
  representations for multi-speaker multi-style text-to-speech,''
\newblock in {\em 2021 IEEE International Conference on Acoustics, Speech and
  Signal Processing (ICASSP)}. IEEE, 2021.

\bibitem{Chou2019VC}
Ju~Chieh Chou and Hung~Yi Lee,
\newblock ``{One-shot voice conversion by separating speaker and content
  representations with instance normalization},''
\newblock in {\em Proceedings of the Annual Conference of the International
  Speech Communication Association, INTERSPEECH}, 2019, vol. 2019-Septe, pp.
  664--668.

\bibitem{T242021M2VoC}
Jie Wang, Yuren You, Feng Liu, Deyi Tuo, Shiyin Kang, Zhiyong Wu, and Helen
  Meng,
\newblock ``The huya multi-speaker and multi-style speech synthesis system for
  m2voc challenge 2020,''
\newblock in {\em 2021 IEEE International Conference on Acoustics, Speech and
  Signal Processing (ICASSP)}. IEEE, 2021.

\bibitem{T032021M2VoC}
Tao Wang, Ruibo Fu, Jiangyan Yi, Jianhua Tao, Zhengqi Wen, Chunyu Qiang, and
  Shiming Wang,
\newblock ``Prosody and voice factorization for few-shot speaker adaptation in
  the challenge m2voc 2021,''
\newblock in {\em 2021 IEEE International Conference on Acoustics, Speech and
  Signal Processing (ICASSP)}. IEEE, 2021.

\bibitem{devlin2018bert}
Jacob Devlin, Ming-Wei Chang, Kenton Lee, and Kristina Toutanova,
\newblock ``Bert: Pre-training of deep bidirectional transformers for language
  understanding,''
\newblock {\em arXiv preprint arXiv:1810.04805}, 2018.

\bibitem{T152021M2VoC}
Zengqiang Shang, Haozhe Zhang, Ziyi Chen, Bolin Zhou, and Pengyuan Zhang,
\newblock ``The thinkit system for icassp2021 m2voc challenge,''
\newblock in {\em 2021 IEEE International Conference on Acoustics, Speech and
  Signal Processing (ICASSP)}. IEEE, 2021.

\bibitem{T132021M2VoC}
Wei Song, Xin Yuan, Zhengchen Zhang, Chao Zhang, Youzheng Wu, Xiaodong He, and
  Bowen Zhou,
\newblock ``Dian: duration informed auto-regressive network for voice
  cloning,''
\newblock in {\em 2021 IEEE International Conference on Acoustics, Speech and
  Signal Processing (ICASSP)}. IEEE, 2021.

\end{thebibliography}

\end{document}